\documentclass{aa}
\usepackage{graphics,amssymb}
\input{psfig}
\begin{document}

\thesaurus{06(02.03.3; 02.20.1; 06.05.01; 06.09.01; 06.18.2)}

\headnote{Research Note}

\title{Shear turbulence beneath the solar tachocline}
\author{E. Schatzman\inst{1}, J.-P. Zahn\inst{1} and P. Morel\inst{2}}
\institute{DASGAL, UMR CNRS 8633, Observatoire de Paris, Section de
Meudon,
92195 Meudon CEDEX, France. \and
D\'epartement Cassini, UMR CNRS 6529, Observatoire de la C\^ote 
d'Azur, BP 4229, 06304 Nice CEDEX 4, France.
}

\offprints{E. Schatzman}
\mail{Evry.Schatzman@obspm.fr}

\date{Received date / Accepted date}
\maketitle

\begin{abstract}
Helioseismic inversions of the Sun's internal angular velocity profile 
show that the rotation changes from differential in latitude
in the convection zone to almost uniform in the radiative 
region below. The transition occurs in a thin layer,
the tachocline,  which is the seat of strong shear in the vertical
direction. In this Note we examine whether this rotation profile can
lead to shear turbulence at the top of the radiation zone.
By using the standard solar
model, we show that such turbulence can be generated only in a narrow 
region 
$0.695\,R_\odot\la r <0.713\,R_\odot$ at the equator and even
in narrower layers at higher latitudes.
We conclude that the turbulence generated by this vertical shear is
unlikely to play a significant role in the transport of matter
and angular momentum, and that other mechanisms must be invoked to 
achieve this.
\end{abstract}

\section{Introduction}
Helioseismology has revealed that the rotation regime in the
Sun changes abruptly from latitude dependent
in the convection zone to almost uniform in the
radiation zone below. The top of the radiation zone
is thus a region of strong vertical shear, and 
for this reason that layer has been called the tachocline 
(Spiegel \& Zahn 1992). Acoustic sounding has shown also
that some mild macroscopic mixing occurs in this region, to
smoothen the composition gradient which, with microscopic
diffusion only, would be somewhat steeper. Moreover, the
Li depletion observed in solar-type stars can only be
explained by some type of macroscopic transport between the convection
zone and the depth where this fragile element is destroyed.

Among the possible mechanisms which may produce such mixing,
a plausible one is the shear instability induced by the vertical rotation
profile. In the present Note we shall verify whether this instability is 
able to play an effective role in the observed mixing.

\section{The solar rotation profile}
Through the inversion of LOWL frequency-splitting data spanning 2 years,
Charbonneau et al. (\cite{ctst98}) have
derived the following internal solar rotation rate for $r>0.5R_\odot$:
\begin{equation}\label{eq:rot}
\Omega(r,\theta)=\Omega_{\rm C}+\frac12\left[1+{\rm erf}
\left(\frac{r-r_{\rm C}}{w}\right)\right]
\left(\Omega_{\rm S}(\theta)-\Omega_{\rm C}\right).
\end{equation}
The observed latidinal differential rotation $\Omega_{\rm S}$
is expressed as:
\[\Omega_{\rm S}(\theta)=\Omega_{\rm 
eq}+c_1\cos^2\theta+c_2\cos^4\theta.\]
The parameters values are:
$\Omega_{\rm C}=2\pi\times432.8$\,nHz, $\Omega_{\rm 
eq}=2\pi\times460.7$\,nHz,
$c_1=-62.69$\,nHz, $c_2=-67.13$\,nHz; $r_{\rm C}=0.713\,R_\odot$ 
is the radius
at the bottom of the solar convective zone and 
$w=0.025\,R_\odot$
the tachocline thickness. Similar expressions can be found in the 
litterature
e.g. Li \& Wilson (\cite{lw98}), Corbard et al.~(\cite{cbpm98}),
Antia et al. (\cite{antia98}).
The fact that $\Omega(r,\theta)$ is almost constant at greater depth
is in conflict with earlier predictions of a faster rotating core
(Pinsonneault et al.~\cite{pins89};
Zahn \cite{z92}), and it rules out that angular momemtum be
transported vertically through shear-induced turbulence.
But this possibility still exists in the shear flow of the tachocline.

\section{Conditions for vertical shear instability}
\subsection{Richardson criterion}
In a stellar radiation zone, the stable stratification of entropy
tends to inhibit any instability arising from a vertical shear $dV/dz$,
where $V$ is the amplitude of the horizontal velocity and $z$ the
vertical coordinate. In the adiabatic limit, i.e. when
radiative losses are negligible, the shear instability
is suppressed whenever:
\begin{equation}\label{eq:richad}
{N^2 \leq \left({d V \over dz}\right)^2} {\rm Ri_c}.
\end{equation}
The strength of the stratification is measured by  
 the Brunt-V\"ais\"al\"a frequency $N$:
\begin{equation}\label{eq:BV}
N^2 = { g \delta \over H_P} \left(\nabla_{\rm ad} - \nabla \right),
\end{equation}
with the usual notations for the gravity $g$, the pressure scale height $H_P$,
the logarithmic temperature gradients $\nabla = \partial \ln T / 
\partial \ln P$ and $\delta = - (\partial \ln \rho / \partial \ln T)_P$,
$P$, $\rho$ and $T$ being pressure, density and temperature.
The critical Richardson number ${\rm Ri_c}$ is of order unity for typical 
flow profiles; in the following we shall take ${\rm Ri_c} = 1/4$.

However shear turbulence may still arise, and be sustained, provided that 
sufficient
heat is lost by the turbulent eddies to lower their buyoancy.
Whether this occurs is determined by  the P\'eclet
number characterizing these eddies:
\[{\rm P_e}=\frac{vl}K;\]
in this expression
 $v$ and $l$ are the velocity and the size of turbulent elements
and $K$ the thermal diffusivity:
\[K=\frac{16\sigma}3\frac{\Gamma_1-
1}{\Gamma_1}\frac{T^4}{\kappa\rho P},\]
$\sigma$ is the Stefan constant,
$\kappa$ the Rosseland mean opacity and $\Gamma_1$ the first 
adiabatic exponent.

When ${\rm P_e} \ll 1$ the Richardson criterion takes the modified form
(Dudis \cite{d74}; Zahn \cite{z74}):
\begin{equation}\label{eq:richmod}
N^2<\left(\frac{dV}{dz}\right)^2\frac{{\rm Ri_c}}{\rm P_e}
\ \Longleftrightarrow vl<\left(\frac{dV}{dz}\right)^2\frac{{\rm Ri_c}K}{N^2} ,
\end{equation}
from which one infers what are the largest turbulent scales that 
can survive in a stratified shear flow.

\subsection{Critical Reynolds number}
However turbulence will be maintained only if the turnover rate of the eddies
is faster than their viscous decay rate:
\begin{equation}\label{eq:rey}
{v \over \ell} \lesssim {\nu \over \ell^2} 
\Longleftrightarrow \nu {\rm Re_c} \leq v \ell ,
\end{equation}
$\nu$ is the viscosity.
This critical Reynolds number $\rm Re_c$ is smaller than
the classical one which governs the onset of instability in a
shear flow whose velocity varies by $U$ between 
two boundaries separated by the distance $L$,
namely $U L / \nu$.  The reason is of course that
$v < U$ and $\ell < L$.

For the value of this $\rm Re_c$ we turn to an experiment
performed by Stillinger et al.~(\cite{shva83}), who measured
the size of the turbulent motions downstream a flow traversing 
a grid. The size of the smallest turbulent eddies was found to be
$15.4 \, \ell_K$, where $\ell_k= (\nu^3/\varepsilon)^{1/4}$ is the classical 
Kolmogorov
length, with $\varepsilon$ being the energy injection rate per unit
mass. In the inertial cascade the velocities scale as 
$v^3 = \varepsilon \ell$, and therefore:
\begin{equation}\label{eq:casc}
v \ell = \varepsilon^{1/3} \ell^{4/3} \geq \varepsilon^{1/3} 
(15.4 \, \ell_K)^{4/3} = (15.4)^{4/3} \nu,
\end{equation}
from which we draw the critical Reynolds number
in that experiment:
$\rm Re_c \approx 40$,
a value we shall adopt for the present purpose.

The conditions (\ref{eq:richmod}) and (\ref{eq:rey}) can be simultaneously
fulfilled when:
\begin{equation}\label{eq:inst}
\left(\frac{dV}{dz}\right)^2>N^2\frac{\nu}K\frac{{\rm Re_c}}{{\rm 
Ri_c}}.
\end{equation}

\section{The turbulent region in the Sun}
\begin{figure}
\centerline{
\psfig{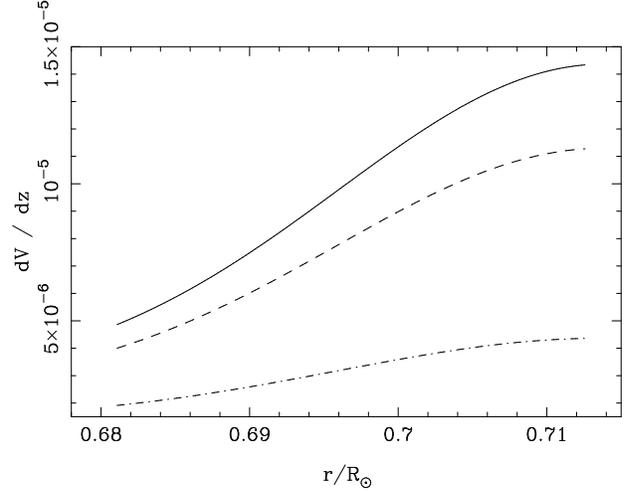}
}
\caption{
The shear rate $dV/dz$ with respect to radius beneath the 
convection zone
for the colatitudes
$\theta=90\degr$ (full), $\theta=60\degr$ (dashed)
and $\theta=30\degr$ (dash-dot-dash).
}\label{fig:dV}
\end{figure}

Let us now examine where the instability condition (\ref{eq:inst}) is
fulfilled is the Sun.
According to (\ref{eq:rot}), the maximum shear rate at the
top of the solar radiation zone amounts to
\[\frac{dV}{dz}\Longrightarrow r\sin\theta\frac{d\Omega}{dr}\approx 9.10^{-
6}\,\rm{s}^{-1},\]
as illustrated in Fig.~(\ref{fig:dV}).

It is convenient to introduce the non-dimensioned parameter
\[\xi\equiv\left(\frac{dV}{dz}\right)^2\frac K{N^2\nu};\]
according to Eq.~(\ref{eq:inst}), the turbulent shear instability will occur if:
\begin{equation}\label{eq:10}
\xi\ga\xi_{\rm limit}\equiv\frac{{\rm Re_c}}{{\rm 
Ri_c}}\approx 160.
\end{equation}
Note that $\xi_{\rm limit}$, the ratio of Reynolds and Richardson
critical numbers, is rather uncertain, in particular because ${\rm Ri_c}$
depends on the vertical flow profile.

We have calculated $\xi$ as a function of radius and colatitude in
the radiative zone beneath the tachocline of a standard solar model,
in order to delimit the regions where the turbulent instability is 
established.

The standard solar model is
computed with CESAM code (Morel \cite{m97}); the evolution 
includes the
pre main-sequence phase.
Basically, the physics and calibration parameters are the same as 
in Morel et al. (\cite{mpb00});
it uses OPAL opacities and equation of state,
the microscopic dif\-fusion coefficients of Michaud \& 
Proffitt~(\cite{mp93}) 
and the recently updated thermonuclear reaction rates
of the European compilation NACRE (Angulo et al. \cite{a99}).
The angular velocity beneath the convection zone
is derived from Eq.~\ref{eq:rot}.
The dynamical viscosity is taken as in a fully ionized hydrogen 
plasma.

\begin{figure}
\centerline{
\psfig{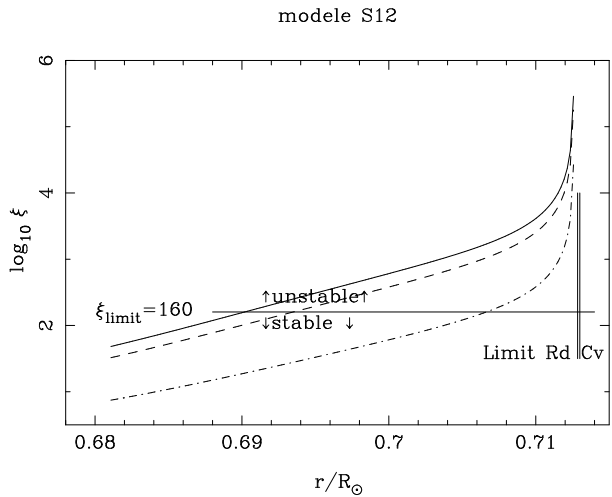}
}
\caption{
For the colatitudes
$\theta=90\degr$ (full), $\theta=60\degr$ (dashed)
and $\theta=30\degr$ (dash-dot-dash),
stable and unstable regions beneath the tachocline located about
at the limit Rd Cv between the convection zone and the radiative 
interior.
}\label{fig:xi}
\end{figure}

\section{Results and discussion}
Figure~\ref{fig:xi} shows $\log_{10}\xi$ as a function of
radius for three colatitude values. According to the Eq.~(\ref{eq:10})
criterion, the turbulent instability
may occur at the equator ($\theta=90\degr$)
in the interval $0.695\,R_\odot \lesssim r <0.713\,R_\odot$;
the turbulent layer is thinner at higher latitudes. 
In other words, the unstable region for the vertical shear instability, 
beneath the
convection zone, measures barely 2 percent of the solar radius
and is localized around the equator.

This instability should not be confused with the shear instability
arising from the differential rotation in latitude, which is not
inhibited by the vertical stratification, and which may play an important
role in smoothing horizontal gradients in composition and angular
velocity (Spiegel \& Zahn \cite{sp92}). But that instability does not
contribute to the vertical transport.

We conclude that another
physical process is needed to transport matter and 
angular momentum beneath the solar tachocline, which would operate more 
efficiently
than the turbulence generated by the vertical $\Omega$-gradient.
A plausible mechanism is the meridional circulation which
probably exists in the tachocline (Brun et al. 1999).
A more powerful process is required also to establish the
almost uniform angular velocity at greater depth,
which could be magnetic torquing 
(Gough \& McIntyre 1998)
or transport of angular
momentum by waves (Schatzman 1996; Kumar et al. 1999).

\begin{acknowledgements}
The authors wish to thank the referee pour his pertinent remarks
and suggestions.
This work has been performed using the computing facilities 
provided by the OCA program
``Simulations Interactives et Visualisation en Astronomie et 
M\'ecanique 
(SIVAM)''.
\end{acknowledgements}

\end{document}